\documentclass[9pt]{article}
\usepackage{spconf,amsmath,graphicx}
\usepackage{hyperref}
\usepackage[fontsize=9pt]{scrextend}

\usepackage{mathtools}
\usepackage{subcaption}
\usepackage{cite} 
\usepackage{booktabs}
\usepackage{enumitem}

\usepackage{amssymb}
\newcommand{\SED}{PSDS}

\newtheorem{Definition}{Definition}

\newcommand\blfootnote[1]{%
  \begingroup
  \renewcommand\thefootnote{}\footnote{#1}%
  \addtocounter{footnote}{-1}%
  \endgroup
}

\title{A Framework for the Robust Evaluation of Sound Event Detection}
\name{\c{C}a\u{g}da\c{s}~Bilen,
Giacomo~Ferroni, 
Francesco~Tuveri,
Juan~Azcarreta and
Sacha~Krstulovi\'{c}}
\address{Audio Analytic - AA Labs,$\:\:$
2 Quayside,
Cambridge, CB5 8AB, UK\\ Email: \{firstname.lastname\}@audioanalytic.com}

\begin{document}
%
\maketitle
\begin{abstract}
This work defines a new framework for performance evaluation of polyphonic sound event detection (SED) systems, which overcomes the limitations of the conventional collar-based event decisions, event F-scores and event error rates. The proposed framework introduces a definition of event detection that is more robust against labelling subjectivity. It also resorts to polyphonic receiver operating characteristic (ROC) curves to deliver more global insight into system performance than F1-scores, and proposes a reduction of these curves into a single polyphonic sound detection score (PSDS), which allows system comparison independently from operating points (OPs). The presented method also delivers better insight into data biases and classification stability across sound classes. Furthermore, it can be tuned to varying applications in order to match a variety of user experience requirements. The benefits of the proposed approach are demonstrated by re-evaluating the baseline and two of the top-performing systems from DCASE 2019 Task 4. 
\end{abstract}
\section{Introduction}
\label{sec:intro}
Sound event detection (SED) is the task of automatically detecting sound events from an audio stream. This benefits many applications such as smart home, smart speakers, headphones, mobile devices, etc.~\cite{DCASE:web, AA:smarthome2018, AA:hearables2019}. Compared to ASR, where sounds always involve vocal tracts and structured language models, sound events are produced by a wide variety of processes happening at quasi random times: this makes SED a challenging problem. SED has received increasing research interest from both academia and industry, as evidenced by the growing number of participants to the {\em Detection and Classification of Acoustic Scenes and Events} (DCASE) data challenge since its start in 2013~\cite{DCASE:2013, DCASE:2016, DCASE:2017, DCASE:2018, DCASE:2019}.

Comparing multiple systems in evaluation campaigns requires agreeing upon a relevant performance criterion. Event-wise and segment-wise error rates have been proposed in~\cite{Giannoulis:2013,Mesaros:2016} and deployed in recent DCASE editions~\cite{DCASE:2016, DCASE:2017, DCASE:2018, DCASE:2019} as a step forward from former frame-based metrics~\cite{DCASE:2013}. However, their current form still ignores the following critical issues:

\textit{Dependency on the operating point:} the same system with different decision thresholds may receive different rankings of performance under the same metric. In other words, such metrics conflate the evaluation of sound event modelling with the evaluation of operating point tuning~\cite{Krstulovic:2018}. This issue has been well studied in signal detection theory, in particular in binary classification, keyword spotting and speaker recognition~\cite{KWSEval2006, Brummer:2006, Brummer:2007}, where ROC curves~\cite{Davis:2006}, Detection Error Trade-off (DET) curves~\cite{Martin1997TheDC} or the area under curve (AUC) metric~\cite{McClish1989} are being used to evaluate a given system globally across a range of operating points. However, this practice has not been as widely adopted into the SED community.

\textit{Definition of sound events:} the event-based metrics defined in~\cite{Giannoulis:2013,Mesaros:2016} rely on collars, which are constraints on the start and end times of the detected events relative to the labelled ground truth events~\cite{DCASE:2016, DCASE:2017, DCASE:2018, DCASE:2019}. The use of collars inherently puts significant emphasis on start and end times of sound events, whereas these timings may prove subjective across human labellers. Therefore, to be robust, an evaluation criterion should allow sufficient room for interpretation of the temporal structure of both the ground truth and the detection timings. 
In this regard, \cite{Tatbul:2018} proposed to decide on true positives (TPs) and false positives (FPs) for anomaly detection on time-series data by relying on percentages of intersection between ground truth and detected events, but this practice has not been applied to SED tasks.

\textit{Prior probabilities, false-positives and cross-triggers in multi-class systems:} cross-triggers (CTs) are the subset of false positives which match another labelled class of the multi-class system. Distinguishing the behavior of CTs from the raw number of FPs can provide insight into data biases, particularly so for sound classes which are acoustically similar. Indeed, multi-class evaluation datasets may become biased in the sense that the amount of data required for reliable evaluation of the TPs for certain target classes may become inconsistent with field priors. For example, window glass breaking rarely happens in practice, yet the reliable evaluation of glass breaking TPs requires a large number of positive class samples, which may in turn artificially increase the FP counts for other percussive classes. Thus, accounting for CTs helps analysing whether FPs are resulting from data bias rather than acoustic modelling defect.

\blfootnote{\textcopyright 20XX IEEE.  Personal use of this material is permitted.  Permission from IEEE must be obtained for all other uses, in any current or future media, including reprinting/republishing this material for advertising or promotional purposes, creating new collective works, for resale or redistribution to servers or lists, or reuse of any copyrighted component of this work in other works.}
In this paper, we propose more robust ways of defining the TPs and FPs, themselves feeding into new definitions of event-based ROC curve and AUC, called Polyphonic Sound Detection ROC (PSD-ROC) curve and Polyphonic Sound Detection Score (PSDS). Section~\ref{sec:background} formally defines the SED task and conventional evaluation approaches while discussing their limitations. The detailed definition of the proposed approach is given in Section~\ref{sec:proposed}, and experimental results demonstrating the benefits of evaluating SED systems with PSD-ROC curves and the PSDS are given in Section~\ref{sec:experiments}. Lastly, conclusions are presented in Section~\ref{sec:conclusions}.

\section{Background}
\label{sec:background}
\subsection{Definition of the Sound Event Detection task}

\begin{figure}[t]
  \centering
  \centerline{\includegraphics[width=\linewidth]{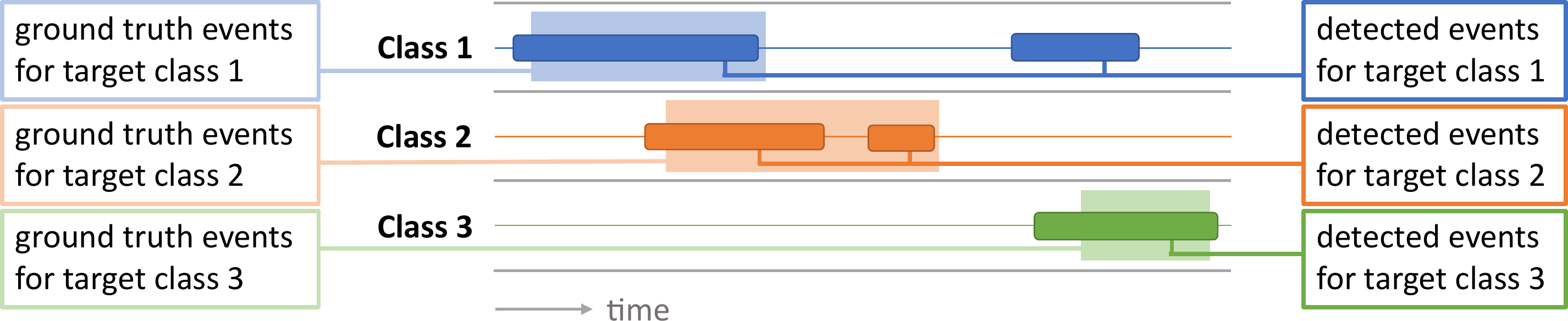}}%
\caption{3-class detection task: transparent boxes represent ground truth labels, solid rounded boxes represent sound detections. }
\label{fig:sed}
\vspace{-4mm}
\end{figure}

In~\cite{Mesaros:2016}, polyphonic sound event detection is defined as the detection of sound events from multiple classes, where it is possible for sound events to occur simultaneously (Figure~\ref{fig:sed}).
Using set theory notations, the SED evaluation task can be formalized as follows:
\begin{Definition}[Event-Based SED Evaluation Task]
\label{def:eval1}
With
\begin{itemize} 
    \item $\mathcal{C}$ being a set of sound classes,
    \item $\mathcal{Y} = \bigcup_{c \in \mathcal{C}}\mathcal{Y}_{c}$ being a dataset which is the union of subsets of ground truth labels for each class $c \in \mathcal{C}$, defined as
    \linebreak $\mathcal{Y}_{c} = \{y_i = (t_{s,i}, t_{e,i}, c_{i}) \colon c_{i} = c\}$, where each ground truth label $y_i$ is defined by its class $c_i$, start time $t_{s,i}$ and end time $t_{e,i}$,
    \item $\mathcal{X}^* = \bigcup_{c \in \mathcal{C}}\mathcal{X}^*_{c}$ being a set of detections which is the union of subsets of detections  for each class $c \in \mathcal{C}$, defined as
    \linebreak $\mathcal{X}^*_{c} = \{x_{j} = (t_{s,j}, t_{e,j}, c_{j}) : c_j = c\}$, where each detection $x_j$ is defined by its class $c_j$, start time $t_{s,j}$ and end time $t_{e,j}$, and where the starred notation $()^*$ indicates dependency on operating point parameters $\mathcal{\tau}_{c}$,
\end{itemize}
the SED evaluation task is defined as measuring the performance of a system which outputs $\mathcal{X}^*$ given $\mathcal{Y}$.
\end{Definition}
The formulation proposed in this paper considers the event-based evaluation of polyphonic systems. However, segment-based evaluation such as in~\cite{Giannoulis:2013,Mesaros:2016}, as well as the evaluation of monophonic systems, can be derived as special cases.

Importantly, SED systems under evaluation are emitting sets of detections \emph{given operating point (OP) parameters $\mathcal{\tau}_{c}, \forall c \in \mathcal{C}$}. Generally, the role of $\mathcal{\tau}_{c}$ is to adjust the permissiveness of the SED system. For example, for SED systems which emit classification scores such as class probabilities, $\mathcal{\tau}_{c}$ may be a set of class-dependent thresholds, where higher thresholds result in making the system more conservative, i.e., emitting the sound detections in which the system is more confident, whereas lower thresholds make the system more permissive by letting more detections through. Various optimisation strategies may be adopted here: some systems may choose to optimize frame classification before factoring frame decisions into event detections, whereas other systems may form event scores before optimising for event-level thresholds. The evaluation approach proposed in this paper aims to cover all SED systems whose operating point can be changed, regardless of their granularity.

\subsection{Limitations of conventional collar-based evaluation}
\label{sec:collar}
The notion of collars introduced in~\cite{Giannoulis:2013,Mesaros:2016} and deployed in DCASE~\cite{DCASE:2013, DCASE:2016, DCASE:2018, DCASE:2019} considers that a ground truth event for class $c$, $y_{i} \in \mathcal{Y}_c$ is correctly detected as a TP if the following condition is satisfied:
\begin{align}
    \exists x_{j} \in \mathcal{X}^*_{c} & \text{ such that } & (t_{s,i} - t_{c}) \leq t_{s,j} \leq (t_{s,i} + t_{c}) \nonumber \\
     & \text{ AND } & (t_{e,i} - \bar{t}_{c}) \leq t_{e,j} \leq (t_{e,i} + \bar{t}_{c})
\end{align}
where $t_{c}$ is a predefined collar duration, and $\bar{t}_{c}$ is the maximum between the collar duration and a predefined ratio of the ground truth duration, i.e. $\bar{t}_{c}\triangleq \max(t_{c}, \rho_c (t_{e,i} - t_{s,i}))$ where $\rho_c$ indicates the predefined ratio. In some applications, where end times are considered difficult to label or irrelevant to the task, the collar around the end time $t_{e,j}$ may be omitted. 

However, collars introduce a limitation which could be hindering system evaluation. Indeed, field application of SED often meets cases where a sound can be reasonably labelled in more than one way due to subjectivity in the interpretation of the temporal structure of sound events. E.g, a dog barking repeatedly could be as reasonably interpreted by human listeners as a single dog barking event, as it could be interpreted as several separate dog bark events, and exhaustive labelling specs turn out difficult to define and enforce in practice. Using collars has the effect of forcing either one or the other of these interpretations to yield a classification error, whereas it would be more desirable for SED evaluations to become robust by design against such reasonable ground truth labelling variability.


\section{Proposed evaluation framework}
\label{sec:proposed}
\subsection{A more robust definition of TPs, FPs and CTs for SED tasks}
\begin{figure}[t]
    \centering
    \begin{subfigure}[b]{1\linewidth}
        \includegraphics[width=\linewidth]{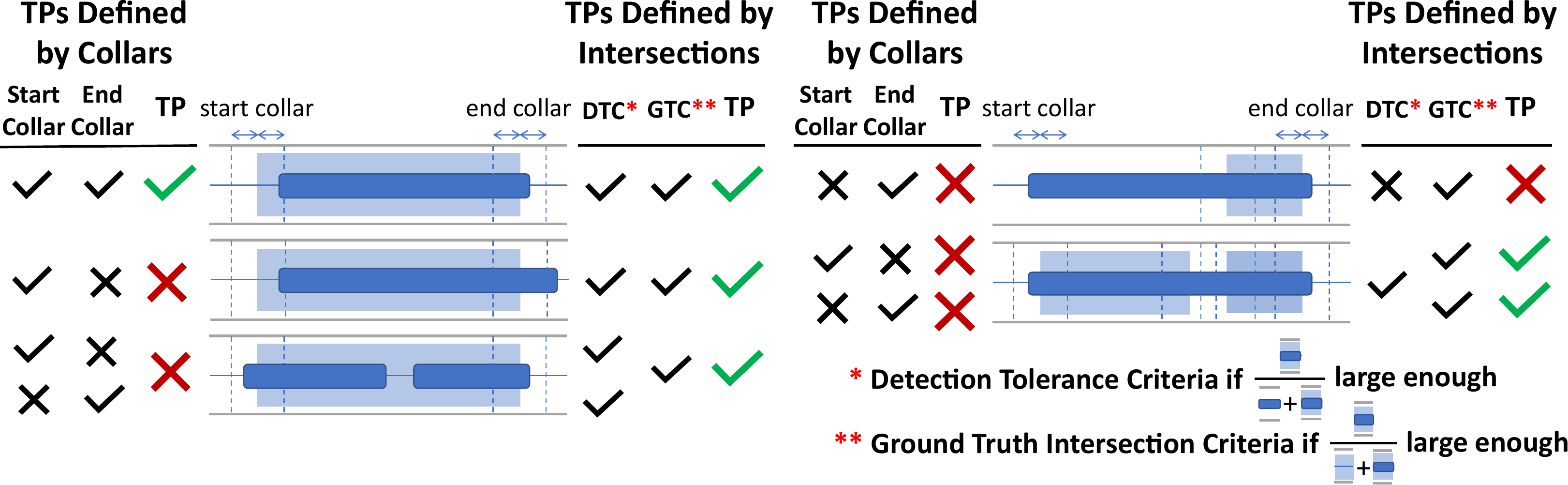}
        \caption{TP decisions made by collars (left) vs. \emph{DTC}/\emph{GTC} (right).}
        \label{fig:res_TP}
    \end{subfigure}
    ~ 
      
    \begin{subfigure}[b]{1\linewidth}
        \includegraphics[width=\textwidth]{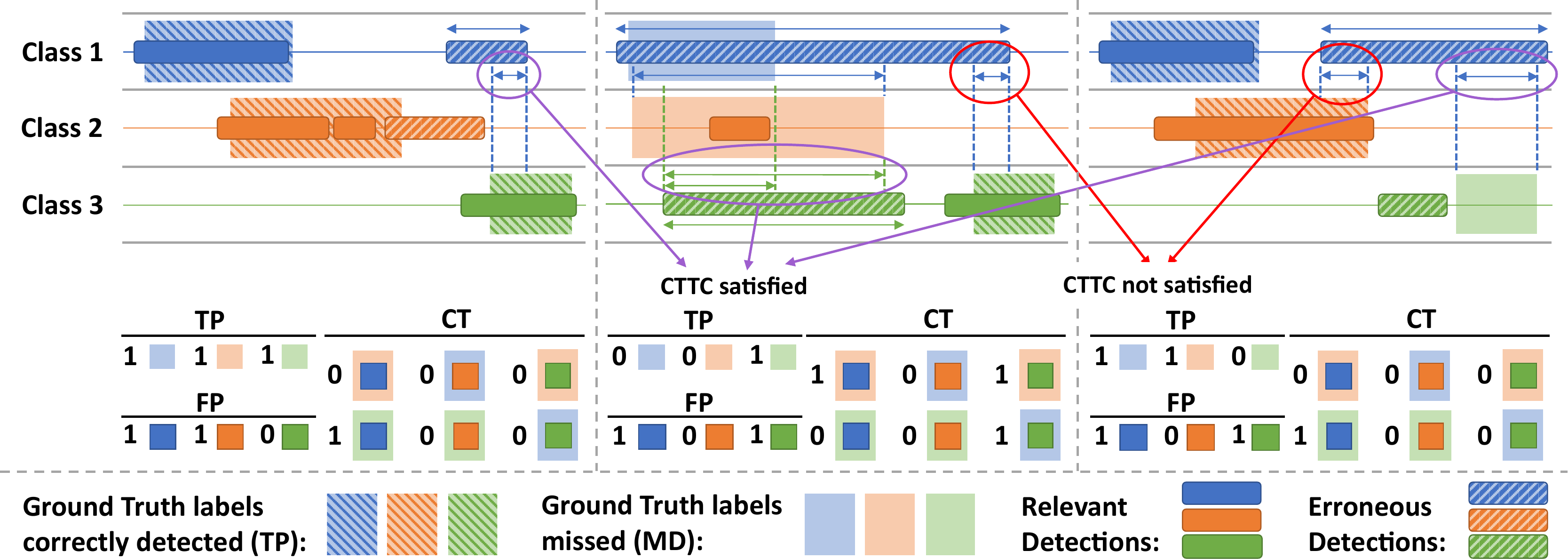}
        ~
        \caption{TPs, FPs and CTs in the proposed method for a 3-class problem. \emph{CTTC} effects are pointed at by the arrows.}
        \label{fig:res_FP}
    \end{subfigure}
    \vspace{-4mm}
    \caption{Visualization of TP, FP and CT counts for \emph{DTC}, \emph{GTC} and \emph{CTTC}, given SED ground truth and detected events. \label{fig:res}}
\vspace{-4mm}
\end{figure}

The proposed evaluation framework is formulated to achieve robustness against (a) operating point variations, (b) labelling subjectivity and (c) possible biases in the evaluation dataset. This starts by redefining TPs and FPs through the combination of three performance criteria defined below and illustrated in Figures~\ref{fig:res_TP} and \ref{fig:res_FP}:

\begin{Definition}[Detection Tolerance Criterion - DTC]
\label{def:RelDet} 
Given a set of ground truth labels $\mathcal{Y}$, the DTC filters a set of detections $\mathcal{X}^*$ to create the set of relevant detections $\mathcal{X}^*_{\text{DTC}}= \bigcup_{c \in \mathcal{C}}\mathcal{X}^*_{\text{DTC},c} \subset \mathcal{X}^*$, based on a detection tolerance parameter $0 \leq \rho_{\text{DTC}} \leq 1$ such that:
\begin{align}
    \mathcal{X}^*_{\text{DTC},c} \triangleq \Big\{ x_j \in \mathcal{X}^*_{c} \colon
      \frac{\sum_{y_i \in \mathcal{Y}_{c}} t_{y_i \cap x_j}}{(t_{e,j} - t_{s,j})} \geq \rho_{\text{DTC}} \:\Big|\: \mathcal{Y}_c,\rho_{\text{DTC}} \Big\}
\end{align}
where $t_{y_i \cap x_j}$ represents the duration of the intersection of the ground truth labels $y_i$ and detected events $x_j$. As a corollary to the above definition, the set of FPs is defined as
$\overline{\mathcal{X}}^*_{\text{DTC}} = \bigcup_{c \in \mathcal{C}}\overline{\mathcal{X}}^*_{\text{DTC},c}$ with $\overline{\mathcal{X}}^*_{\text{DTC},c} \triangleq \mathcal{X}^*_c \setminus \mathcal{X}^*_{\text{DTC},c}$ and the number of FPs per class is defined as the cardinality of this set:
$N^*_{\text{FP},c} = \Big| \overline{\mathcal{X}}^*_{\text{DTC},c} \Big|$.
\end{Definition}

\begin{Definition}[Ground Truth intersection Criterion - GTC]
\label{def:TP}
The GTC creates the set of correctly detected ground truth events, $\mathcal{Y}_{\text{TP}} = \bigcup_{c \in \mathcal{C}} \mathcal{Y}_{\text{TP}, c}  \subset \mathcal{Y}$, given the set of relevant detections $\mathcal{X}^*_{\text{DTC},c} \forall c \in \mathcal{C}$ and a ground truth tolerance parameter $0 \leq \rho_{\text{GTC}} \leq 1$ such that:
\begin{align}
\label{eq:first}
    \mathcal{Y}^*_{\text{GTC},c}  \triangleq 
    \Big\{y_i \in \mathcal{Y}_c \colon 
   \frac{\sum_{x_j \in \mathcal{X}^*_{\text{DTC},c}} t_{y_i \cap x_j}}{(t_{e,i} - t_{s,i})} \geq \rho_{\text{GTC}} \:\Big|\: \mathcal{X}^*_{\text{DTC},c},\rho_{\text{GTC}} \Big\}
\end{align}
The number of TPs is then defined as the cardinality of the GTC sets across all classes: $N^*_{\text{TP}} = \sum_{c \in \mathcal{C}}N^*_{\text{TP}, c}$ where $N^*_{\text{TP}, c} = \big| \mathcal{Y}^*_{\text{GTC},c } \big|$ .
\end{Definition}

Similarly to~\cite{Tatbul:2018}, \emph{DTC} and \emph{GTC} measure percentages of intersection between ground truth labels and detected events. Our approach differs in the way intersections are thresholded to count the TPs/FPs before calculating the final performance figure. Having observed on our industrial data sets that inter-labeller disagreements happen mostly at event boundaries, e.g., where sound events fade away or are made of units whose boundary is subject to interpretation, an approach based on intersection tolerances rather than boundary collars is inherently more robust, as shown in figure~\ref{fig:res_TP}.

Coming back to Definition~\ref{def:RelDet}, some of the FPs could be due to specific data biases which may surface as confusions between target sound classes. Hence, the \emph{cross-trigger tolerance criterion (CTTC)} is introduced in Definition~\ref{def:CT} to allow counting the CTs separately, as illustrated in Figure~\ref{fig:res_FP}.

\begin{Definition}[Cross-Trigger Tolerance Criterion - CTTC]
\label{def:CT}
Given a set of ground truth events $\mathcal{Y}$, the cross-trigger tolerance criterion counts the CTs as $N^*_{\text{CT}} = \sum_{c \in \mathcal{C}} \sum_{\substack{\hat{c} \in \mathcal{C}\\ \hat{c} \neq c}} N^*_{\text{CT},c,\hat{c}}$, given FP set $\overline{\mathcal{X}}^*_{\text{DTC}}$  and a cross-trigger tolerance parameter $0 \leq \rho_{\text{CTTC}} \leq 1$ such that:
\begin{align}
    &N^*_{\text{CT},c,\hat{c}} = \sum_{x_j \in \overline{\mathcal{X}}^*_{\text{DTC},c}}  \sum_{\mathcal{Y}_{\hat{c}}: \substack{\hat{c} \in \mathcal{C}\\ \hat{c} \neq c}} \left[ \frac{\sum_{y_i \in \mathcal{Y}_{\hat{c}}}t_{y_i \cap x_j}}{(t_{e,j} - t_{s,j})} \geq \rho_{\text{CTTC}} \right]
\end{align}
where sets $\mathcal{Y}_{\hat{c}}$ select the ground truth for each class $\hat{c} \neq c$. Notation $\left[\cdot\right]$ represents the Iverson bracket, which denotes 1 when the enclosed condition is satisfied and 0 otherwise.
\end{Definition}




\subsection{Performance metrics relevant to user experience}
\label{sec:perf}

In field applications, user experience (UX) may be heavily influenced by the frequency of FPs~\cite{Krstulovic:2018}. As a result, and given Definitions~\ref{def:TP}-\ref{def:CT} and tolerance parameters $\rho_{\text{DTC}}$, $\rho_{\text{GTC}}$ and $\rho_{\text{CTTC}}$, the \emph{TP ratio}, \emph{FP rate} and \emph{CT rate} may be computed as:
\begin{align}
    \textbf{TP Ratio:}\; r^*_{\text{TP},c} = \frac{N^*_{\text{TP},c}}{\big| \mathcal{Y}_c \big|}\quad\quad\quad\quad
    \textbf{FP Rate:}\; R^*_{\text{FP},c} = \frac{N^*_{\text{FP},c}}{T_{\mathcal{Y}}}  \nonumber \\
    \textbf{CT Rate:}\; R^*_{\text{CT},c, \hat{c}} =  \frac{N^*_{\text{CT},c,\hat{c}}}{\sum_{y_i \in \mathcal{Y}_{\hat{c}}} (t_{e,i} - t_{s,i})}\quad\quad\quad
\end{align}
where $T_{\mathcal{Y}}$ is the total duration of dataset $\mathcal{Y}$. Thus, TP performance is measured as a proportion of detected events, whereas FP and CT performances are rates per unit of time, consistently with keyword spotting evaluation practice~\cite{KWSEval2006}. $R^*_{\text{FP},c}$ relates to the total duration of the dataset, whereas $R^*_{\text{CT},c, \hat{c}}$ is only relevant to target class labels. 

Moreover, cross-triggers against identified sound classes may trigger more negative user experience than less identifiable FPs~\cite{Krstulovic:2018}, thus justifying the definition of the \emph{effective FP rate (eFPR)} as:
\begin{align}
\label{eq:mean_ct}
    \textbf{eFPR:}\; e^*_{c} \triangleq R^*_{\text{FP},c} +\alpha_{\text{CT}} \dfrac{1}{\vert \mathcal{C}\vert - 1} \sum_{\substack{\hat{c} \in \mathcal{C}\\ \hat{c} \neq c}}R^*_{\text{CT},c,\hat{c}}
\end{align}
where weighting parameter $\alpha_{\text{CT}}$ represents the cost of CTs on user experience in the SED application under evaluation.

System behaviour may not guarantee that the sets of operating points $\mathcal{O}_c = \big\{(r^*_{\text{TP},c}, e^*_{c}), \forall \mathcal{\tau}_{c} \big\}$ will form convex or monotonous class-dependent ROC curves: there may exist a set $\bar{\mathcal{O}}_c$ of OPs which are yielding a lower TP ratio at a higher eFPR than other OPs
\begin{align}
\bar{\mathcal{O}}_c = \big\{
(\bar{r}^*_{\text{TP},c}, \bar{e}^*_{c}) \; : \;
\exists \; (r^*_{\text{TP},c}, e^*_{c}) \; \text{with}
\begin{array}{c}
     r^*_{\text{TP},c} > \bar{r}^*_{\text{TP},c}    \\
     e^*_{c} \leq \bar{e}^*_{c}
    \end{array}
\big\}
\end{align}
However, these would never be chosen as practical OPs for a product if adjacent points provide a better trade-off, thus they are discarded to 
form the largest set $\widehat{\mathcal{O}}_c$ of best case operation points for each class:
    $\widehat{\mathcal{O}}_c = \mathcal{O}_c \setminus \bar{\mathcal{O}}_c$.

At this stage, varying $\mathcal{\tau}_{c}$ corresponds to moving monotonically across $(\hat{r}^*_{\text{TP},c}, \hat{e}^*_{c})$ points in $\widehat{\mathcal{O}}_c$, so the operating point dependency $()^*$ can be dropped and the discrete best case envelopes $\widehat{\mathcal{O}}_c$ can be interpolated into continuous ROC curves $r_{\text{TP},c}(e)$ by holding the preceding value $r_{\text{TP},c}(e) = \hat{r}^*_{\text{TP},c}(\hat{e}^*_{c}) , \forall e > \hat{e}^*_{c}$ (``staircase-type'' interpolation). Thus, a set of class-dependent ROC curves $r_{\text{TP},c}(e)$ becomes available for averaging into a single polyphonic ROC curve.
%

However, the {\em stability of performance across classes} is of interest for evaluation: systems with much smaller variations in the TP ratio across classes may be preferable due to having better performance for the worst performing (or most difficult) class. For this reason, the \emph{effective TP ratio (eTPR)} is defined using both the mean and the standard deviation of TP ratios across classes, such that:
\begin{align}
\label{eq:mean_std}
    \mu_{\text{TP}} =  \dfrac{1}{\vert \mathcal{C}\vert} \sum_{c \in \mathcal{C}}r_{\text{TP},c} \quad
    \sigma_{\text{TP}} = \sqrt{\dfrac{1}{\vert \mathcal{C}\vert} \sum_{c \in \mathcal{C}}\left(r_{\text{TP},c}-\mu_{\text{TP}}\right)^2}\\
\label{eq:etpr}
    \textbf{eTPR:}\qquad r(e) \triangleq \mu_{\text{TP}}(e) - \alpha_{ST} * \sigma_{\text{TP}}(e)\qquad\qquad
\end{align}
where parameter $\alpha_{\text{ST}}$ adjusts the {\em cost of instability across classes} for the SED task under evaluation. In principle, the mean and standard deviation in~\eqref{eq:mean_std} could also be computed with different weights for each class, thus enabling further adaptability of the method.

At this point, calculation of the (eFPR, eTPR) coordinates has reduced a set of class-dependent ROC curves into one single polyphonic sound detection (PSD) ROC curve $r(e)$, which remains to be summarised into a single number. Given equations~\ref{eq:first}-\ref{eq:etpr} and Definitions~\ref{def:eval1}-\ref{def:CT}, the normalized area under the PSD ROC curve is thus computed to define the system's PSD score (PSDS) as follows:
\begin{Definition}[Polyphonic Sound Detection Score]
\label{def:sedmetric}
Given a dataset with a set of ground truth events, $\mathcal{Y}$, and the set of evaluation parameters, $(\rho_{\text{DTC}}, \rho_{\text{GTC}}, \rho_{\text{CTTC}}, \alpha_{\text{CT}}, \alpha_{\text{ST}})$, a SED system's \SED{} is:
\begin{align}
    \text{PSDS} \triangleq \frac{1}{e_{\text{max}}}\int_{0}^{e_{\text{max}}} r(e) \, de
\end{align}
where $e_{\text{max}}$ is the maximum \emph{eFPR} value of interest for the SED application under evaluation.
\end{Definition}

\section{Experimental results}
\label{sec:experiments}

To demonstrate the benefits of the proposed \SED{} metric, three publicly available SED systems submitted to DCASE challenge 2019 Task 4 are evaluated under both DCASE's collar-based metric~\cite{Giannoulis:2013,Mesaros:2016} and the proposed \SED{} metric\footnote{A public and open source implementation of the proposed PSDS metric is available from: \href{https://github.com/audioanalytic/psds_eval}{\url{https://github.com/audioanalytic/psds\_eval}}}. These systems are: the baseline system~\cite{Serizel2018}, $1^{\text{st}}$ ranking~\cite{Lin2019} and $4^{\text{th}}$ ranking~\cite{Cances2019} systems of Task~4, renamed System 1, 2 and 3 respectively. Three experiments demonstrate the advantages of the proposed method in defining TPs and FPs in a more robust way~ (\ref{sub:robustness}), achieving independence from the operating point~(\ref{sub:psds}) and incorporating inter-class confusions and classification stability into the final evaluation~(\ref{sub:adapting}). For all three experiments, the validation set of DCASE 2019 Task 4 is used as the test set for evaluation, since ground truth labels for the official test set were not released at the time of writing.

\subsection{Robustness of a DTC/GTC-based \texorpdfstring{$F_1$}{F1}-score}
\label{sub:robustness}
During DCASE 2019, the three systems were evaluated using the collar-based $F_1$-score~\cite{Giannoulis:2013,Mesaros:2016}. This is compared in Table~\ref{tab:F1score_comp} against recomputing the TPs, FPs and corresponding $F_1$-scores using the DTC and GTC with $(\rho_{\text{DTC}}, \rho_{\text{GTC}}) = (0.5, 0.5)$ and $(0.8, 0.8)$. 
\begin{table}[h]
    \centering
    \begin{tabular}{@{} *4l @{}}    \toprule
    \emph{Evaluation method} & \emph{System 1} & \emph{System 2} & \emph{System 3}  \\\midrule
     F1-Score (collars) & 23.70~\% & 42.32~\% & 39.90~\% \\ 
     F1-Score ($\rho_{\text{DTC}}, \rho_{\text{GTC}} = 0.5$) & 61.38~\% & 60.13~\% & 63.24~\% \\ 
     F1-Score ($\rho_{\text{DTC}}, \rho_{\text{GTC}} = 0.8$) & 37.03~\% & 39.51~\% & 43.57~\% \\\bottomrule
     \hline
    \end{tabular}
    \caption{Evaluation of three SED systems using collar-based $F_1$-scores vs DTC/GTC $F_1$-scores. }
    \label{tab:F1score_comp}
\end{table}
After re-interpretation of TPs and FPs, the ranking of the systems changes, and they can be seen to perform very closely with respect to the DTC/GTC F1-score. Indeed, detailed examination of the errors reveals that many FPs for System~1 were due to split detections which are reinstated as TPs using the DTC/GTC approach. The performance measures for the three systems are similarly evened out when increasing the DTC/GTC tolerance parameters to 0.8. This demonstrates how various definitions of TPs and FPs may influence the perception of performance. 

\subsection{Summarising performance into a single \SED{} figure}
\label{sub:psds}
\begin{figure}[t]
    \centering
    \begin{subfigure}[b]{.5\linewidth}
        \includegraphics[width=\linewidth]{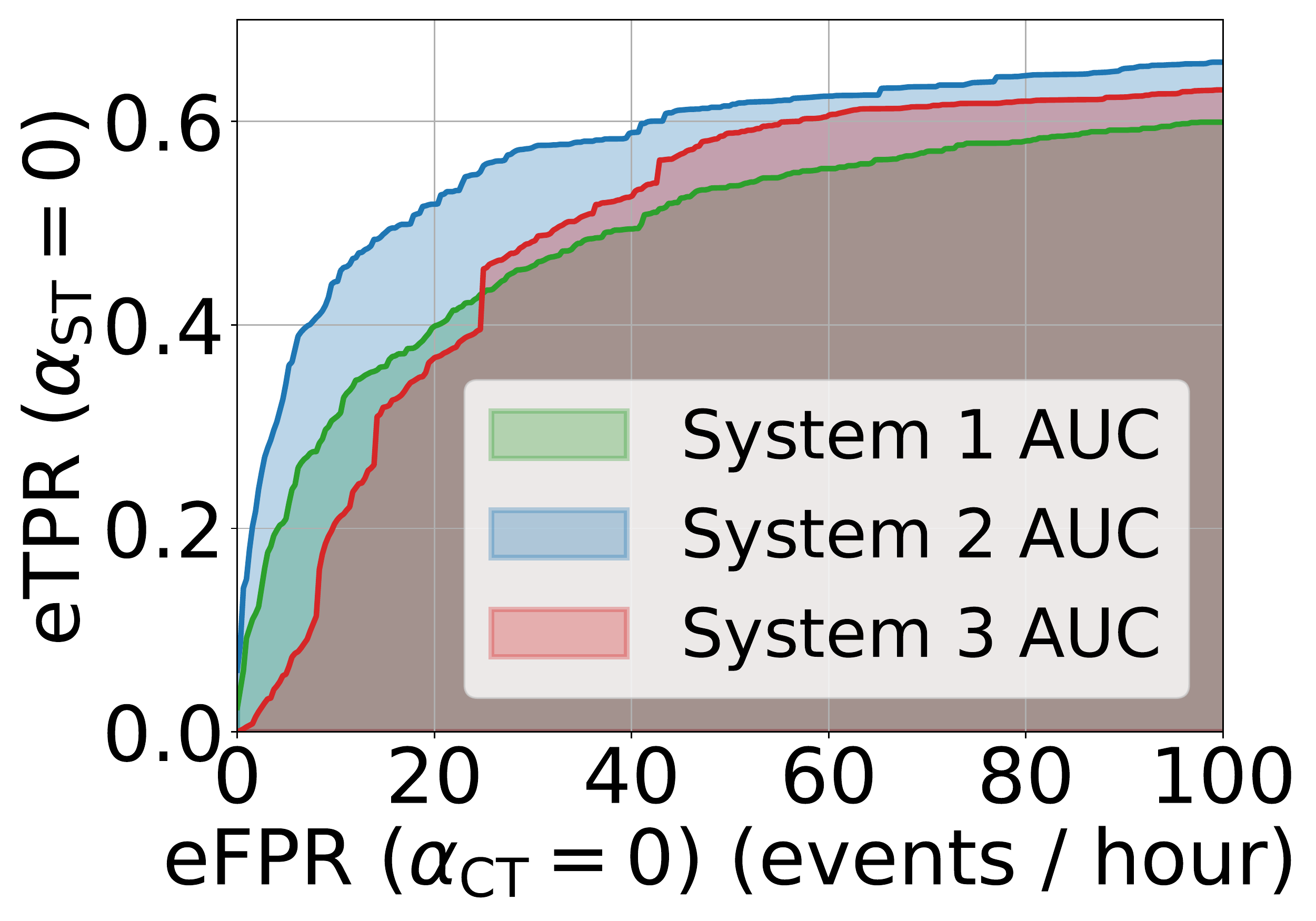}
        \caption{$\alpha_{\text{CT}} = 0$.}
        \label{fig:result1}
    \end{subfigure}%
        \hfill
    \begin{subfigure}[b]{.5\linewidth}
        \includegraphics[width=\linewidth]{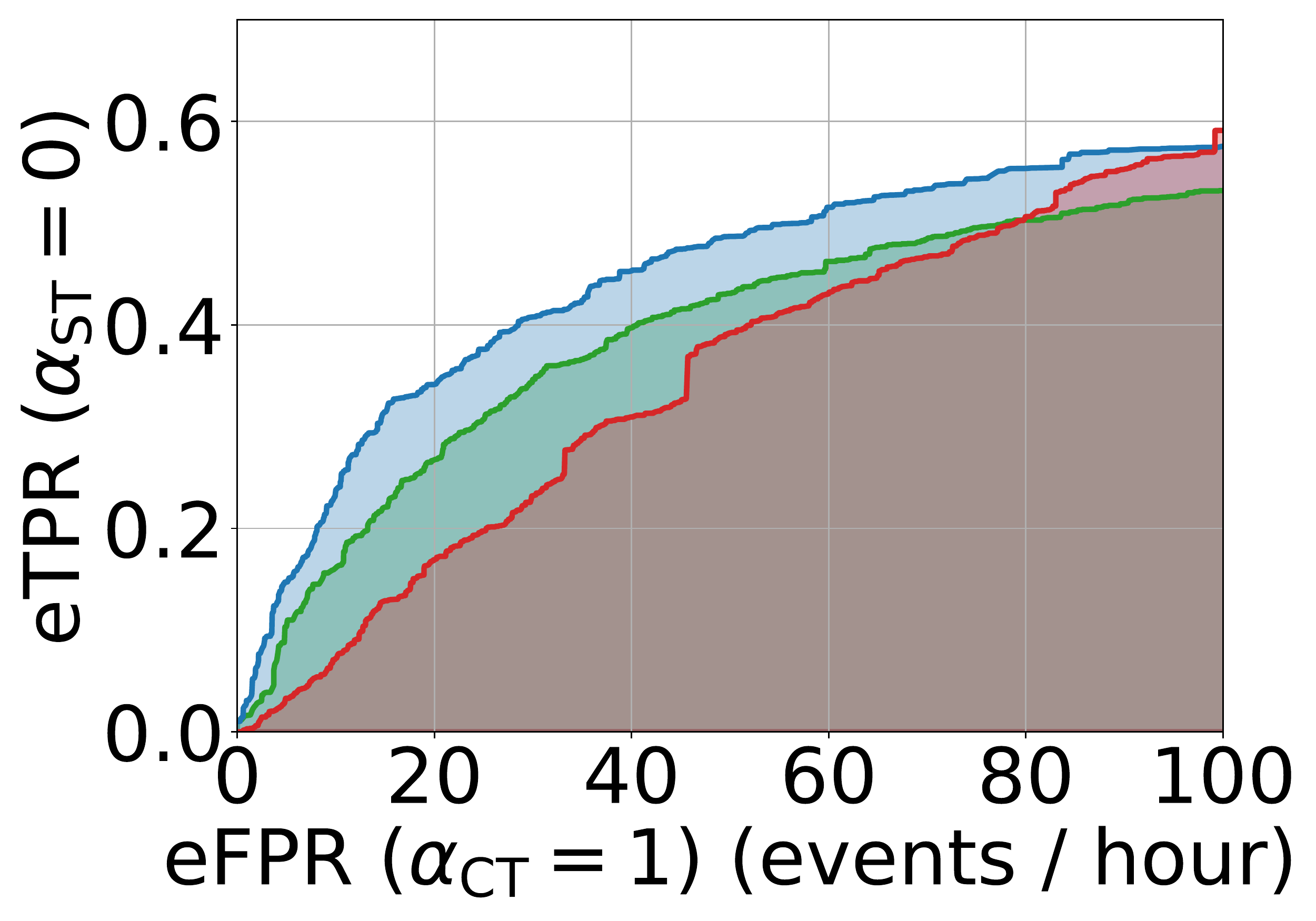}
        \caption{$\alpha_{\text{CT}} = 1$.}
        \label{fig:result2}
    \end{subfigure}
    \caption{The PSD-ROC curves and areas used for computation of the PSD scores given on the first and second rows of Table~\ref{tab:sed_metric_comp}.}
    \label{fig:result_comparison}
\vspace{-4mm}
\end{figure}

The PSDS is more comprehensive than the $F_1$-scores from Table~\ref{tab:F1score_comp}, insofar as it factors in the system performance across a range of OPs, as well as FP rates per unit of time. To test the effects of that on system ranking, the thresholds used by the corresponding authors for each system have been varied\footnote{System~2's implementation has been slightly modified to allow varying its operating point, hence its performance differs slightly from the original submission.} to create the PSD-ROC curves shown in Figure~\ref{fig:result1}, with parameters $(\rho_{\text{DTC}}, \rho_{\text{GTC}}, \rho_{\text{CTTC}}, e_{\text{max}}) = (0.5,0.5,0.3,100)$. The CTs and stability across classes are ignored by setting $\alpha_{\text{CT}}$ and $\alpha_{\text{ST}}$ to 0.

The PSD-ROC curves in Figure~\ref{fig:result1} show that System~2 can perform better than the other two systems under all the performance trade-offs, whereas System~1 and System~3 do not have a clear superiority to each other: System~1 could be preferable for very low FP rates whereas System~3 performs better at higher TP ratios. Such insight into the global advantage of System~2 would not be possible to achieve if looking only at $F_1$ scores specialized to single operation points, e.g., from Table~\ref{tab:F1score_comp}. The PSD scores of the three systems are given on the first row of Table~\ref{tab:sed_metric_comp}, and confirm the global advantage of System~2 at evaluation setting $(\alpha_{\text{CT}}, \alpha_{\text{ST}}, e_{\text{max}}) = (0, 0, 100)$.

\subsection{A flexible evaluation criterion}
\label{sub:adapting}
Different systems may show different advantages under evaluation criteria which may vary according to the desired application and related user experience factors. Indeed, Table~\ref{tab:sed_metric_comp} shows that the average scores, the score differences and the ranking of systems may change under varying evaluation criteria. However, the significant reduction of scores observed when $\alpha_{\text{ST}}=1$ indicates that all the systems suffer from unstable classification performance across classes. On the other hand, System~1 has an advantage over System~3 in low FP rate/eFPR settings, thus confirming what was observed in Figure~\ref{fig:result_comparison}. Similarly, System~1 may have an advantage over System~3 for applications where the cross-triggers have a particular impact on UX, i.e. $\alpha_{\text{CT}}=1$, at high eFPR settings, as can be seen in Figure~\ref{fig:result2}.
\begin{table}[h]
    \centering
    \begin{tabular}{@{} *4l @{}}    \toprule
    \emph{PSDS parameters} & \emph{System 1} & \emph{System 2} & \emph{System 3}  \\\midrule
     $(\alpha_{\text{CT}}, \alpha_{\text{ST}}, e_{\text{max}}) = (0, 0, 100)$ & 0.486 & \textbf{0.573} & \underline{0.493} \\ 
     $(\alpha_{\text{CT}}, \alpha_{\text{ST}}, e_{\text{max}}) = (1, 0, 100)$ & \underline{0.385} & \textbf{0.442} & 0.342\\ 
     $(\alpha_{\text{CT}}, \alpha_{\text{ST}}, e_{\text{max}}) = (0, 1, 100)$ & \underline{0.336} & \textbf{0.377} & 0.313\\ 
     $(\alpha_{\text{CT}}, \alpha_{\text{ST}}, e_{\text{max}}) = (0, 0, 50)$ & \underline{0.398} & \textbf{0.507} & 0.372 \\\bottomrule
     \hline
    \end{tabular}
    \caption{The PSD scores under different evaluation settings. Best score is in \textbf{bold}, second best is \underline{underlined}. Bigger is better.}
    \label{tab:sed_metric_comp}
\vspace{-4mm}
\end{table}

\section{Conclusions}
\label{sec:conclusions}
This paper presents a new evaluation framework for SED systems, whose benefits are demonstrated by re-evaluating the baseline and two of the top-ranking systems from DCASE 2019 Task 4, using the new metric over the same task and data. Firstly, the new definition of TPs and FPs is shown to remove a major limitation of collars because it uses intersection criteria rather than boundaries when matching against ground truth, thus yielding performance measurements which are more robust against labelling subjectivity. Secondly, the proposed PSD-ROC curve and related PSD score provide more comprehensive insight than $F_1$ scores into system performance, because they span ranges of operating points rather than single system settings, and factor in the FP rates per unit of time. Lastly, the proposed method can be adapted to the varying needs of different applications and different user experience requirements, by adjustment of evaluation parameters. Insofar as evaluation metrics are essential to the discovery of progress across successive generations of a technology or across systems compared within public data challenges such as DCASE, it is hoped that the PSDS metric presented in this paper will contribute to steering SED research in useful directions.

\clearpage
\vfill\pagebreak

\bibliographystyle{IEEEbib}
\bibliography{strings,refs}

\end{document}